\documentclass[11pt,preprint]{emulateapj}
\usepackage{natbib}
\usepackage{color}
\usepackage{epsfig}
\usepackage{comment}
\usepackage{amsmath}
\usepackage{graphicx}
\shorttitle{Imaging the AGN Torus in Cygnus A}

\shortauthors{Carilli, Perley, Dhawan, Perley}
 
\begin{document}

\title{Imaging the AGN Torus in Cygnus A}

%\correspondingauthor{}
%\email{ccarilli@nrao.edu}

\author{
C. L. Carilli\altaffilmark{1,2},
R.A. Perley\altaffilmark{1},
V. Dhawan\altaffilmark{1},
D.A. Perley\altaffilmark{3}
}

\altaffiltext{1}{National Radio Astronomy Observatory, P. O. Box 0,Socorro, NM 87
801, USA; ccarilli@nrao.edu, ORCID: 0000-0001-6647-3861}                                                     
\altaffiltext{2}{Astrophysics Group, Cavendish Laboratory, JJ Thomson Avenue, Cam
bridge CB3 0HE, UK}                                                              
\altaffiltext{3}{Astrophysics Research Institute,
John Moores University, 146 Brownlow Hill,
Liverpool L3 5RF, UK
}

\begin{abstract}

We present the first direct imaging of what may be the thick torus in
the active galactic nucleus (AGN) of the archetype powerful radio
galaxy Cygnus A, using the Jansky Very Large Array (VLA) at 18 GHz to
48 GHz, with a resolution down to 45 mas. Such a torus has long been a
key component of AGN models, but direct imaging on the relevant
physical scales in sources of extreme (quasar-like) luminosities,
remains scarce. An elongated structure, perpendicular to the radio
jets and centered on the core, is well resolved, with a full length of
$0.48"$ (528 pc), and a full width of $0.26"$ (286 pc).  The radio
emission spectrum is consistent with optically thin free-free
emission. We present a toy model of a flaring torus, with a
half-opening angle for the poloidal region of $62^o$.  The radio jets
are oriented along the poles. The observations require a clumpy gas
distribution, with the free-free emission dominated by clumps with
densities $\ge 4000$ cm$^{-3}$

\end{abstract}


\section{Introduction}

An active galactic nucleus (AGN) corresponds to the process of
prodigious energy production by gas in-fall onto a supermassive black
hole (SMBH; \citet{hick18, netzer15, ramos17}). In order to explain the
observed  spectral properties of AGN at gamma-ray through radio
wavelengths, a number of structural elements have been hypothesized:
On scales $< 1$ pc, a hot, thin accretion disk is formed immediately
around the SMBH, giving rise to X-ray emission, and to the ultraviolet
'big blue bump'. Highly ionized gas clouds in this inner zone
give rise to broad UV and optical emission lines (line
widths $> 1000$ km s$^{-1}$). On scales of a few pc to ~ 100 pc, gas from
the host galaxy falls into a zone dominated by the gravitational
potential of the SMBH.  Conservation of angular momentum drives the
gas to form a toroidal structure, centered on the SMBH. This torus is
seen predominantly through infrared emission from warm dust. The
poloidal regions of the torus are lower density, and the residual gas
clouds in these regions are ionized by radiation from the AGN, giving
rise to narrow (few hundred km s$^{-1}$), optical emission lines. At
larger radii, beyond the radius where the supermassive black hole
dominates the gravitational potential, the structures then merge with
the large-scale structures of the parent galaxy itself. In some 10\% of
AGN, highly collimated relativistic jets of plasma are generated,
oriented along the polar axis of the torus, and emitting strong radio
synchrotron emission.

While the above picture may be idealized, a structure acting as a
thick obscuring torus is a defining aspect of the AGN unification
model: Type 1 AGN (broad line), correspond to sources where our line
of sight passes down the axis of the polar region, providing an
unobscured view of the hot accretion disk and broad line region. For
Type 2 AGN (narrow line), our line of sight to the broad line region
and hot accretion disk passes through the torus, and hence these
regions are obscured in the optical by dust in the torus
\citep{hick18, urry95, anton84, barv87}.  The obscuring torus is
part of the general unification model of radio loud quasars and radio
galaxies, with the former being the radio loud Type I AGN, and the
latter, Type II. The radio jets propagate along the polar direction of
the torus. Certainly, the rich relationship between Type I and II AGN
is more involved \citep{pad17}, however, the notion of an obscuring
torus remains paramount in the physics and unification of Type I and
II AGN.

An important conclusion regarding AGN obscuring tori is
clumpiness. Both Type I and II AGN show a similar correlation between
the mid-IR emission and hard X-ray emission, implying a large mean
free path for both X-ray and mid-IR photons in the torus.  Hence, the
tori must be clumpy and multi-phase, with dense clouds of
self-shielding dust embedded in a medium of atomic and ionized gas
\citep{tacconi94, nenk08, ramos17}.  Collisions between clumps are
also invoked to increase turbulent pressure, to help provide
vertical support of the torus, along with radiation pressure, and
possibly magnetic pressure \citep{netzer15, krolik88, krolik07}.

Observations with near-IR interferometers, and with the {\it HST}, as well
as high resolution ALMA observations of gas and dust, are starting to
resolve torus-scale regions in growing samples of AGN, such as
Centaurus A, NGC 1068, Circinus, 3C 273, and others \citep{garcia16,
izumi18, burt13, trist09, gall97, gall04, gall16, combes18, aalto17,
iman18, grav18, jaffe96, jaffe04}. However, the samples are, for the
most part, comprised of lower luminosity AGN, and even in the more
extreme cases, such as 3C 273, the black hole masses are still
an order of magnitude smaller than in Cygnus A.

Cygnus A, at z = 0.0562, is ten times closer than the next radio
galaxy of similar radio luminosity\footnote{Radio luminosity $>
10^{45}$ erg s$^{-1}$}.  The nuclear regions in Cygnus A have been
observed extensively at radio through X-ray wavelengths
\citep{cari96}.  The inner few arcseconds is a complex mix of
optically obscuring dust clouds \citep{vest93, lopez14, whys04,
merlo14}, atomic gas seen in narrow line emission \citep{stock94,
taylor03}, HI 21cm absorption toward the inner radio jets, with a
neutral atomic column density $> 10^{23}$ cm$^{-2}$, depending on HI
excitation temperature \citep{struve10}, polarized, broad optical
emission lines due to scattering by dust \citep{anton94, ogle97}, and
a highly absorbed hard X-ray spectrum with a total gas column density
of $\sim 3\times 10^{23}$ cm$^{-2}$ \citep{reynolds15, ueno94}. VLBI
radio observations at 0.05 mas resolution reveal highly collimated
jets originating on scales $\sim 200$ times the Schwarzschild radius
\citep{bocc16}. \citet{tad03}, derive a black hole mass of $2.5\pm 0.7
\times 10^{9}$ M$_\odot$ from {\it HST} and Keck spectroscopy of
Pa-$\alpha$ and [OIII], and conclude that Cygnus A contains an AGN
with a bolometric luminosity of order 10$^{46}$ erg s$^{-1}$,
comparable to high redshift quasars \citep{runnoe12}. This AGN is
highly obscured in the optical due to dust along our line of sight,
with $A_v > 50$ magnitudes, based on near-IR spectroscopy
\citep{iman00}.  Studies of the mid- to far-IR spectral and
polarization properties have led to a model of a clumpy, dusty torus
obscuring the AGN in Cygnus A, with a radius of at least 130 pc,
although these conclusions are based on spatially integrated
properties; these observations did not have the spatial resolution to
resolve the torus, and hence are partially contaminated by emission
from the radio core-jet \citep{privon12, lopez18}.

While substantial evidence exists for an obscured AGN with a
quasar-like luminosity in the center of Cygnus A, to date, no direct
imaging observations have been made of the torus at a resolution and
sensitivity sufficient to resolve the torus structures on the relevant
scales of 50 pc to 100 pc. In this letter, we present VLA observations
of the nuclear regions of Cygnus A down to 45 mas
resolution\footnote{At the distance of Cygnus A, $1" = 1.1$mas}.  We
detect a prominent structure perpendicular to the radio jet axis,
centered on the core. The emission spectrum of this structure is
consistent with free-free emission. We model this as emission from the
ionized gas associated with the thick, clumpy torus.

\section{Observations, Calibration, and Imaging} \label{sec:obs}

Cygnus A was observed for 5 hours on October 21, 2016, and on April
28, 2018, with the VLA in its A configuration. Observations were made
in five frequency bands, including the 18 GHz to 26 GHz band, the 30
to 38 GHz band, and the 40 to 48 GHz band, using the wide band (8 GHz)
correlator set-up in full polarization\footnote{We also present
spectral points for the radio core from lower frequencies observed as
part our larger Cygnus A VLA program. These data will be presented in
full elsewhere (Perley et al. in prep).}.

Editing and calibration of the data utilized the AIPS processing
system. The flux density scale was set using
coefficients for 3C48 for 2016 observations, and 3C286
for 2018 observations \citep{perley13a}. The accuracy of the flux scale,
including errors associated with transfer of the flux density scale to
the target objects, is estimated to be $< 5\%$. Bandpass and delay
calibration utilized the nearby point-source J2007+4029, which was in
turn utilized for amplitude and phase calibration of the target source
Cygnus A. Polarization calibration utilized J2007+4029 to determine
the leakage terms, and both 3C48 and 3C286 to set the linear
polarization position angle \citep{perley13b}.

Self-calibration was employed starting with a point
source model for the nucleus of Cygnus A. The
nucleus dominates the visibilities above 20 GHz, at the
resolution of the VLA A-configuration, where the lobes and hot spots
are highly attenuated by the primary beam and highly resolved by the
interferometer. Further self-calibration using the resulting
clean component model resulted in a final dynamic range $\sim 10^5$.

The visibility data, with the optimum radio core flux subtracted (see
the appendix), were then ported to CASA, where imaging was performed
using a multifrequency synthesis CLEAN, with Briggs weighting with
robust = -0.5. The imaging was performed separately on 2GHz wide
sub-bands within the bands, for a total of 12 frequency images from 19
GHz to 47 GHz. The primary beam of the VLA has a FWHM of $1.9'$ at 23
GHz, $1.3'$ at 34 GHz and $1.0'$ at 44 GHz. The full double lobed
radio source in Cygnus A has an extent of $2'$, hence the hot spots at
the extremities of the lobes, and outer lobes, which dominate the
luminosity of the source, are highly down-weighted by the primary
beam.  Still, the shortest visibility spacings showed substantial
emission from large scale structure, and a uv-minimum of 200
k$\lambda$ was employed to avoid artefacts from large scale
structure. At 34 GHz and 43 GHz, a uv-taper was employed to obtain a
CLEAN Gaussian restoring beam of FWHM = 45 mas. At 22 GHz, the
restoring clean beam had FWHM = 67 mas, and images in the 34 GHz and 44 GHz
bands were also made at 67 mas resolution.

\section{Results} \label{sec:results}

Figure 1 shows the resulting images in each high frequency band, and
Figure 2 shows the summed image of the three bands, all after
subtraction of the strong nuclear point source (see appendix).  The
resolution is 45mas in the 34 GHz and 44 GHz bands, and 67 mas in the
22 GHz band and in the summed image. The jets are seen ranging from
$0.1”$ to $1.0"$ distance from the radio core, oriented roughly $14^o$
north of west. The recently discovered radio transient is detected
$0.4"$ southwest of the nucleus \citep{perley17}. Two knots of
emission are seen within 50 mas of the core, along the jet directions,
in these core-subtracted images. These knots may correspond to the
inner radio jets seen on small scales \citep{cari94, sor96, krich98},
although, based on analysis of the core subtraction process (see the
appendix), we restrict our quantitative analysis to radii further than
about the FHWM of the lower resolution image ($> 70$ mas), from the
core.

The focus of this paper is the structure seen extending perpendicular
to the jet direction, symmetrically to the north and south of the
core. The structure is very similar in the high resolution images at
Ka and Q band. The full lengths of the major and minor axes at the
4$\sigma$ surface brightness level in the summed image are 480 mas and
260 mas, respectively. In the analysis below, we adopt the term
'torus' for this structure, for ease of reference.

Figure 3 shows a slice in surface brightness along the torus in the 32
GHz image, through the core position and perpendiculat to the jet. We
do not consider the structure in the inner $\pm 50$ mas reliable, due
to unknowns in the core subtraction process (see appendix).
Beyond 50 mas distance from the core, the torus is reasonably well
characterized by Gaussian wings with a FWHM = 138 mas, out to about
125 mas radius. Beyond that, the torus shows a broader skirt, again
extending to about 240 mas radius, before it is lost in the noise.

We imaged the polarized emission from the nuclear regions of Cygnus
A. A faint (4.5 mJy) point source is present in the polarized
intensity image at the position of the core, implying a fractional
polarization at 35GHz of 0.3\%. We consider this an upper limit, given
it is at the limit of the polarization calibration capabilities of the
VLA \citep{perley13b}. No polarized emission is seen anywhere else in
the region, down to a level of 50 $\mu$Jy beam$^{-1}$ at 35 GHz, and
70 $\mu$Jy beam$^{-1}$ at 43 GHz.  The upper limit to the fractional
polarization at the positions of the torus considered in the spectra
below is about 10\% (1$\sigma$). The limits for the brighter inner
halo, and bright jet knot 150 mas northwest of the core, are a few
percent.

\section{Analysis}
\label{sec:analysis}

\subsection{Spectrum and $T_B$}

Figure 4 shows spectra at different positions along the torus. Figure
4a is a spectrum of the radio core itself, made at resolutions
ranging from the 45 mas at high frequency, to about $0.8"$ at the
lowest frequency. At the lower resolutions and frequencies, we
estimate, based on the structures seen in Fig. 1, that the jets and
disk contribute as much as 20\% to the total flux density in the lower
frequency points in Figure 4a. However, the uncertainty in this
estimate is substantial, and we make no correction for this
contamination at lower frequency to the data points in Figure 4a.
Figures 4b and c show spectra in the torus, at positions 100 mas north
and south of the core, along the disk major axis, at 67 mas
resolution.

We plot two power-law model spectra in Figs. 4b and 4c, with indices
of -0.1 and -0.5, respectively. The former is appropriate for
optically thin free-free emission. The latter would be the spectrum of
Thomson scattered radiation from the radio nucleus (ie. same spectrum
as the nucleus, see below). The free-free model fits the data
reasonably, with a cumulative probability based on the reduced
$\chi^2$ in both cases $\sim 70\%$. The Thomson scattering model is
clearly much too steep relative to the data, with a formal probability
$< 10^{-7}$. We note that synchrotron radiation from spatially diffuse
emission typically has an even steeper spectrum (index $\sim
-0.75$). Likewise, thermal emission from warm dust would have a
sharply rising (Rayleigh-Jeans) spectrum at these frequencies.

The surface brightnesses in the 34 GHz image at the two torus
positions shown in the spectra are 0.447 mJy beam$^{-1}$ and 0.496 mJy
beam$^{-1}$. We adopt the mean value of 0.47 mJy beam$^{-1}$ in the
following analysis. The implied brightness temperature in the torus at
100 mas from the core is $T_B = 240$ K. If the emission were optically
thin free-free emission, the Emission Measure is then: $EM = 12.0
\times T_B \times T_e^{0.35} \times \nu^{2.1}$ pc cm$^{-6}$ $= 1.2
\times 10^8$ pc cm$^{-6}$, where $T_e$ is the electron temperature in
K (assumed to be $10^4$ K, see below), and $\nu$ is the observing
frequency, in GHz \citep{condon18}.  As a point of comparison,
observations of free-free absorption toward the counter-jet in the low
luminosity radio AGN in the nearby galaxy, NGC 1275, determined an
emission measure of $5 \times 10^8$ pc cm$^{-6}$ on scales of a few
parsecs, which they attribute to the AGN accretion disk or torus
\citep{walker00}.
 
The spectrum of the radio core shows a sharp turn-over at frequencies
below 10 GHz. The model in Figure 4a is a power-law spectrum, with the
power-law index of -0.5 set by fitting to the 34 GHz and 44 GHz data,
plus free-free absorption at lower frequency using an opacity set by
the emission measure value measured in the torus above: $\tau_{ff} = 3.3
\times 10^{-7} \times \nu^{-2.1} \times EM$ \citep{condon18}, with
$\nu$ in GHz. Curiously, the location in frequency of the predicted
spectral turnover based on the EM of the torus,
$\nu_{\tau_{ff} = 1} = 5.7$ GHz, is
close to the observed spectral turnover. However, we cannot conclude
that free-free absorption is the dominant cause for the turnover, for
the following reason.

VLBI imaging of the core-jet of Cygnus A at 22 GHz and 43 GHz down to
resolutions of 0.1 mas show 90\% of the flux density at 43 GHz in the
VLA spectrum comes from jet and counterjet structures within 3 mas
radius of the nominal radio core. These structures show a variety of
spectral indices between 20 GHz and 43 GHz, ranging from -1 to
+1. The ‘radio core’ is identified as the most inverted (sharply
rising) spectral component, likely due to synchrotron self-absorption
\citep{krich98}. 

We calculate the expected turnover frequency due to synchrotron
self-absorption of the jet components in Cygnus A. We adopt the
outermost jet knot, J7, in the 43 GHz VLBI image, about 2 mas from the
radio core. This component has a flux density of 42 mJy and a size of
0.36 mas, derived from Gaussian-fitting \citep{bocc16}. The implied
equipartition magnetic field is 0.1 G \citep{miley80}. The implied
frequency for the turnover (peak frequency), for synchrotron
self-absorption is given by: $\nu_{max} = 0.051 \times (B^{1/2} \times S
\times \nu^{-\alpha})^{2/(5 - 2\alpha)}$ GHz = 6 GHz \citep{pach70}, where B
is in Gauss, $S$ is the surface brightness in Jy arcsec$^{-2}$ at
observing frequency $\nu$, in GHz, and $\alpha$ is the spectral
index\footnote{ This equation is good to 15\% for $\alpha$ in the
range -0.5 to -0.8 \citep{pach70}}. VLBI knots closer to the core will
have even higher turnover frequencies.

Hence, even for the outer knots on these VLBI scales, the expected
turnover frequency due to synchrotron-self absorption is $\ge 4$ GHz.
The implication is that the spectral turnover is as plausibly due to
synchrotron self-absorption, as due to free-free absorption, or
possibly a combination of these phenomena.

\subsection{Polarization}

We imaged the linear polarized emission from the nuclear regions, and
find no polarized emission from the torus or the jets to limits of a
few to 10\%. Polazation could potentially be used as a diagnostic on
the emission mechanism: Thermal free-free emission is expected to be
unpolarized.  Scattered radiation is naturally polarized, eg. the
scattered optical broad lines in Cygnus A are 16\% polarized
\citep{ogle97}.  Synchrotron radiation from the jet would also likely
be significantly intrinsically polarized $\sim$ tens of percent
\citep{zensus97, bridle84}.

However, Faraday Rotation complicates the analysis.  For instance,
taking a mean density of $n_e = 500$ cm$^{-3}$, a pathlength of $L =
250$ pc, and assuming the magnetic fields in the nuclear regions of
galaxies can be of order a milli-Gauss \citep{mcbride14}, implies a
rotation measure of: $RM = 0.81 \times n_e \times B \times L \sim
10^8$ rad m$^2$, with $n_e$ in cm$^{-3}$, $B$ in Gauss, and $L$ in pc.
This rotation measure would cause 5000 radians of rotation of the
intrinsic position angle for the polarized emission, even at 43 GHz
(7mm). Hence, in practice, any minute gradient in this RM screen
across the synthesized beam , or any mixing of such a screen in the
emitting regions, would depolarize the emission, even at sub-cm
wavelengths. Hence, the lack of polarization from the inner radio jet
likely results from large values of Faraday Depolarization, as the
light propagates through the dense nuclear regions of Cygnus A.

\subsection{Torus Model} 

Overall, while we cannot prove that the emission from this torus in
thermal free-free emission, the flat spectra of the torus regions are
consistent with such a model, and inconsistent with either Thomson
scattering, diffuse synchrotron emission, or thermal emission from
warm dust. Similarly, free-free emission from these regions does not
uniquely imply a torus \citep{gall04}. However, the observed spatial
distribution implies a structure substantially elongated perpendicular
to the radio jets, and symmetrically distributed north and south of
the core, and in the following discussion we adopt a toy model
representation of a torus, for further investigation.

We assume that the dense material is in a cone-like region, with the
polar regions of this torus oriented roughly in the sky plane. The
radio jets propagate in the polar directions. We note that the latest
VLBI analysis of the radio jets on pc-scales suggests a jet
inclination angle with respect to the sky plane of $15.5^o$
\citep{bocc16}. The torus, in this model, is a mixture of dense,
self-shielded dust clouds, enveloped in atomic and ionized gas
\citep{nenk08, netzer15}. A schematic of the model is shown in Figure
5.

The maximum observed radius of the torus is about $R_t \sim 264$ pc,
while the height above the midplane is $H_t \sim 143$ pc. Note that
the ratio of length to width is not a measure of inclination angle.
That would be true only for a thin disk. For a thick torus, regardless
of the mean inclination angle, the AGN must be obscured along our line
of sight, and therefore some part of the torus must cover our line of
sight to the core. Assuming an approximate conical structure, the
observed ratio of radius over the height is roughly the tangent of the
half-opening angle of the poloidal regions of the torus, implies
a half-opening angle of $\theta_t \sim 62^o$.

The mean density can be derived from the emission measure, given a
pathlength through the torus. In this model, the diameter of the torus
is: $L_{diam} = 528$ pc, but we measured the EM at 100 mas from the
core along the major axis, which reduces the pathlength to:
$L_{100mas} = 513$ pc. Using $EM = n_e^2 \times L_{100mas} =
1.2\times 10^8$ pc cm$^{-6}$, then implies a mean density of: $n_e =
490$ cm$^{-3}$.

However, if the torus was filled with a diffuse gas of this mean
density, the following issue arises.  The column density through the
torus to the core is then: $(L_{diam}/2) \times n_e \sim 4 \times
10^{23}$ cm$^{-2}$.  The Thomson scattering optical depth is: $t_e =
\sigma_e \times (L_{diam}/2) \times n_e = 0.26$, where $\sigma_e$ is
the electron Thomson scattering cross section. Removing the poloidal
areas (which cover roughly half the sky seen from the core), means
that $\sim 13\%$ of the light from the radio core will be Thomson
scattered isotropically, i.e., that the torus would have an integrated
radio flux of $\sim 0.13 \times 1.3$ Jy = 170 mJy at 34 GHz. The
integrated flux density for the entire torus region seen in Figure 1
is only 21 mJy at 34 GHz after core subtraction, or about 8 times
less.

The challenge is that, while the spectra of the torus argue for
free-free emission, the process cannot be simply due to a diffuse
ionized medium with the mean density as calculated above, or else the
emission would be dominated by Thomson scattering, and have a
different spectrum.  One solution to this apparent paradox is to
assume that the ionized gas in the torus is clumpy.  The free-free
line emission behaves as $n_e^2$, while Thomson scattering is linear
in $n_e$. The free-free emission will therefore be strongly weighted
toward the densest regions. For instance, a density 8 times higher
than the mean in the line emitting clumps would require a total
pathlength 64 times less to produce the emission measure, and would
decrease the overall optical depth to Thomson scattering by a factor
8.

As discussed in the introduction, models for dusty tori invoke
clumpiness in the dust distribution to explain the relative mid-IR and
X-ray properties \citep{nenk08, ramos17}. It seems natural to extend
the clumpy structure into the neutral and ionized gas distributions as
well. In this case, the free-free emission is predominantly from
regions with densities $\ge 4000$ cm$^{-3}$.  

For comparison, based on their HI 21cm absorption measurements toward
the radio jet in Cygnus A, \citet{struve10} quote a density on similar
scales of $\ge 10^4$ cm$^{-3}$, comparable to our value.  They also
quote a scale height of the torus of 20 pc at a radius of 80
pc. In our simple linear flared torus model, the scale height at 80 pc
radius is a factor two larger.

\citet{taylor03} performed slit spectroscopy with a
spatial resolution of $0.9"$ of the narrow line emission regions of
Cygnus A. While the resolution is clearly not adequate to separate
structures on the scales studied herein, the average results on optical
line ratios are still useful as rough estimates of the temperatures in
the narrow line emitting gas in the nuclear regions. The temperatures
they derive range from 8200 K to 15000 K. For a temperature of $10^4$
K, the implied lower limit to the pressure in the free-free emitting
clumps is $5.5 \times 10^{-9}$ dynes cm$^{-2}$.

For comparison, we consider the minimum pressure in the relativistic
particles and magnetic fields in the inner radio jet. The surface
brightness of the jet as it emerges from the torus region in Figure 1,
at roughly 100 mas from the core, is 1.2 mJy beam$^{-1}$ at 34.5
GHz. Assuming the jet is marginally resolved at 45 mas, the implied
minimum non-thermal pressure is $1.5 \times 10^{-8}$ dynes cm$^{-2}$
\citep{miley80}. Hence, the lower limit to the pressure in the dense
ionized gas is within a factor three of the minimum pressure in the
relativistic jets.

\section{Discussion} \label{sec:disc}

We present the discovery of an elongated, flat spectrum, radio
continuum structure in the nuclear regions of Cygnus A, which we
designate the `torus'. This structure is centered on the radio core,
with a length of $0.48"$ (528 pc), oriented perpendicular to the
jet, and a full width of $0.26"$ (286 pc). The radio spectra of this
region are consistent with flat spectrum thermal free-free emission,
and inconsistent with Thomson scattering of radio emission from the
radio core, diffuse synchrotron emission, and thermal emission from
warm dust. We consider these results in the context of models
for clumpy, multi-phase obscuring tori in AGN. 

Our toy model for the thick torus leads to a half-opening angle for
the poloidal region of $62^o$. For comparison, in the clumpy torus
models of \citet{nenk08}, they calculate that the observed
demographics of Type 2 versus Type 1 AGN, in which the fraction of
Type 2 to Type 1 AGN lies between 50\% and 70\% (see for example
\citet{wilkes13, hick18}), can be explained by
tori with half-opening angles between $45^o$ and $63^o$, depending on
the nature of the torus (sharp-edged or Gaussian-edged).  They do caution
that their models pertain to lower luminosity AGN, with lower mass
black holes, as seen in eg.  Seyfert galaxies.

For the torus size, models for dusty tori in AGN typically adopt an
inner radius set by the sublimation of graphite dust by radiation from
the AGN. This radius is small, of order a parsec, even for an AGN with
$10^{46}$ erg s$^{-1}$ \citep{kish13, burt13, ramos17,netzer15}. Such
a small scale for the inner torus radius is supported by near-IR
interferometric observations, as the near-IR probes the hottest dust,
near the sublimation radius \citep{burt13, netzer15}.

Physical predictions for the outer radius of the torus are more
qualitative, and indeed, the exact location may be ill-defined, as the
torus structures merge with the larger scale structures of the host
galaxy. As an approximate guide for the outer torus radius,
\citet{hick18} suggest that: 'a natural scale for the outer edge of
the torus is the gravitational sphere of influence of the SMBH' (see
also \citet{netzer15}).  \citet{tad03} make a rough estimate of the
stellar mass density in the inner 100 pc of Cygnus A of 10 M$_\odot$
pc$^{-3}$. At this mass density, the mass in stars will exceed that of
the SMBH at a radius of 390 pc. A similar calculation can be done
using equation (3) in \citet{netzer15}, and using the stellar velocity
dispersion of the Cygnus A galaxy of 290 km s$^{-1}$
\citep{thornton99}. This calculation leads to a black hole sphere of
influence radius of 130 pc.  While these are clearly very rough
estimates, it appears the maximum radius of the torus as seen in the
radio image (264 pc), is in the range expected for the radius at which
the gravitational dynamics transitions from being dominated by the
SMBH to dominance by the overall mass distribution in the host
galaxy.

The torus size suggested herein is larger than those seen in lower
luminosity AGN presented in recent studies (see \citet{ramos17} for a
review, and references in Sec. 1). In terms of ionization, the AGN in
Cygnus A is certainly luminous enough to maintain ionization to this
radius. For example, \citet{netzer15} equation (6) implies a radius of
460 pc for the narrow line region in an AGN with a bolometric
luminosity of $\sim 10^{46}$ L$_\odot$. However, such a model also
requires that ionizing AGN photons can leak through the torus out to
these radii. Again, such leakage is a natural consequence of a clumpy
torus. As emphasized by \citet{nenk08}: 'The (clumpy) models can
produce nearly isotropic IR emission together with highly anisotropic
obscuration, as required by observations. Clumpiness implies that the
viewing angle determines an AGN classification only probabilistically:
a source can display Type 1 properties even from directions close to
the equatorial plane.'  Of course, we cannot rule-out contributions to
the ionization by star formation in the inner few hundred parsecs of
Cygnus A \citep{privon12, hoffer12}, nor a contribution to the
dust-obscuration by dust on larger-scales in the host galaxy of Cygnus
A \citep{merlo14}.

The fact that Cygnus A is so anomalously close, given its AGN
luminosity, has clearly enabled the
observation of the possible toroidal structure on scales of tens to a
few hundred parsecs. For comparison, observations of the next closest
radio galaxy of similar radio luminosity, 3C 295 at $z = 0.464$, would
require five times the angular resolution and five times the
sensitivity to perform measurements at similar physical resolution and
brightness sensitivity as we have obtained on Cygnus A. The situation
gets dramatically worse with increasing redshift due to cosmological
surface brightness dimming. The almost edge-on orientation of the torus in
Cygnus A (as implied by the jet orientation angle; \citep{bocc16}),
also facilites these findings.  Previous observations of Cygnus A
lacked the combination of spatial resolution, frequency coverage, and
sensitivity to image such a structure \citep{carilli91,
carilli99}. Only with the new systems of the Jansky Very Large Array,
in particular, the complete frequency coverage up to 50 GHz and the
order of magnitude improvement in continuum sensitivity, have these
observations become possible \citep{perley11}.

It is the fortuitous combination of proximity, luminosity, and
orientation that make Cygnus A such a prime target for future studies
of the possible torus.  Higher resolution, higher sensitivity
observations are required to determine with greater accuracy the
morphology and spectrum of this new structure perpendicular to the
jets in the nucleus of Cygnus A. The current observations argue for
thermal emission from a region that would be consistent with a
multi-phase, clumpy torus, comprised of dense clouds of obscuring dust
immersed in more diffuse, but also clumpy, atomic and ionized gas.

\acknowledgments The National Radio Astronomy Observatory is a
facility of the National Science Foundation operated under cooperative
agreement by Associated Universities, Inc.. Observations were made
under the VLA observing programs: 16B381, 16B396, and 17B200.
We thank C. Ricci for useful comments, and C. Ricci and C. Ramos-Almeida
for reproduction of their figure. We acknowledge support from
Chandra Grant G05-16117B.

\clearpage
\newpage

\appendix

\section{Point Source Subtraction}

Imaging with these data clearly show the outer regions of the torus,
but details within $\sim \pm 50$ mas of the core are complicated by the
restored `clean beam' response of the bright nuclear radio source. To
address this issue, a point source model was subtracted from the
visibilities, varying the strength to minimize residual artefacts.
The nominal 'best fit' subtraction was adjusted to get the flattest
brightness emission profile through the torus along the line
perpendicular to the jets. We then added and subtracted values ranging
from $\pm 4$ mJy of this nominal value, in increments of 1 mJy, 
and investigated the residuals.

Figure A1 shows the resulting images at 31.5 GHz for different levels
of core subtraction.  Besides the rising and falling point source at the core
position by the roughly 1 mJy value expected from the changing model,
we determine the surface brightnesses at 40 mas and 70 mas distance
from the core along the torus axis at 32 GHz.  At 40 mas distance, the
surface brightnesses do change, from 1.6 mJy beam$^{-1}$ to 2.6 mJy
beam$^{-1}$ as the subtracted point source flux density is changed
from +4 mJy to -4 mJy around the adopted 'best' value. However, at
70mas distance, the value remains constant at 1.4 mJy beam$^{-1}$, to
within 3\%.  Hence, we limit our analysis to greater than 70 mas from
the core.

Note that the data were taken on two dates separated by 1.5 years,
between which the nuclear core increased by ~130 mJy. Hence, the core
subtraction process was done separately for each date, and the
resulting databases combined.

The typical brightness of the torus is less than 0.1\% of that of the
nuclear emission at this resolution, a critical question is whether
the torus emission is real, and is not some manifestation of the
calibration process. While arguments based on the similarity of the
structure over a wide frequency range (18 through 48 GHz) are
reassuring, a better argument comes from utilizing the same
calibration and imaging, and source subtraction, methods on a
different source of similar strength.  The nearby phase calibrator,
J2007+4029, has nearly the same flux density and spectrum as that of
the Cygnus A nucleus, and was observed on the same days, at the same
frequencies and over the same hour angle range as Cygnus A. Thus, we
analyzed the calibrator data in exactly the same manner (Fig. A2). We
find that no extended `halo' emission exists on the angular scale of
that found in Cygnus A, to a brightness level less than 1/10 that of
Cygnus A. J2007+4029 does have a one-sided weak jet, extending about
$1.5"$ to the south-west of the nuclear emission. Further evidence of
the reality of the residual emission is the presence of the two
components, located 25 mas from the nucleus along the outer jet axes,
as well as the extension seen on mas-scales by in VLBI imaging
\citep{push17}.

\clearpage
\newpage

\begin{figure}[!t]
\centering 
\epsscale{1.25}
%\plotone{fig1.pdf}
\includegraphics[scale=0.8,angle=-90]{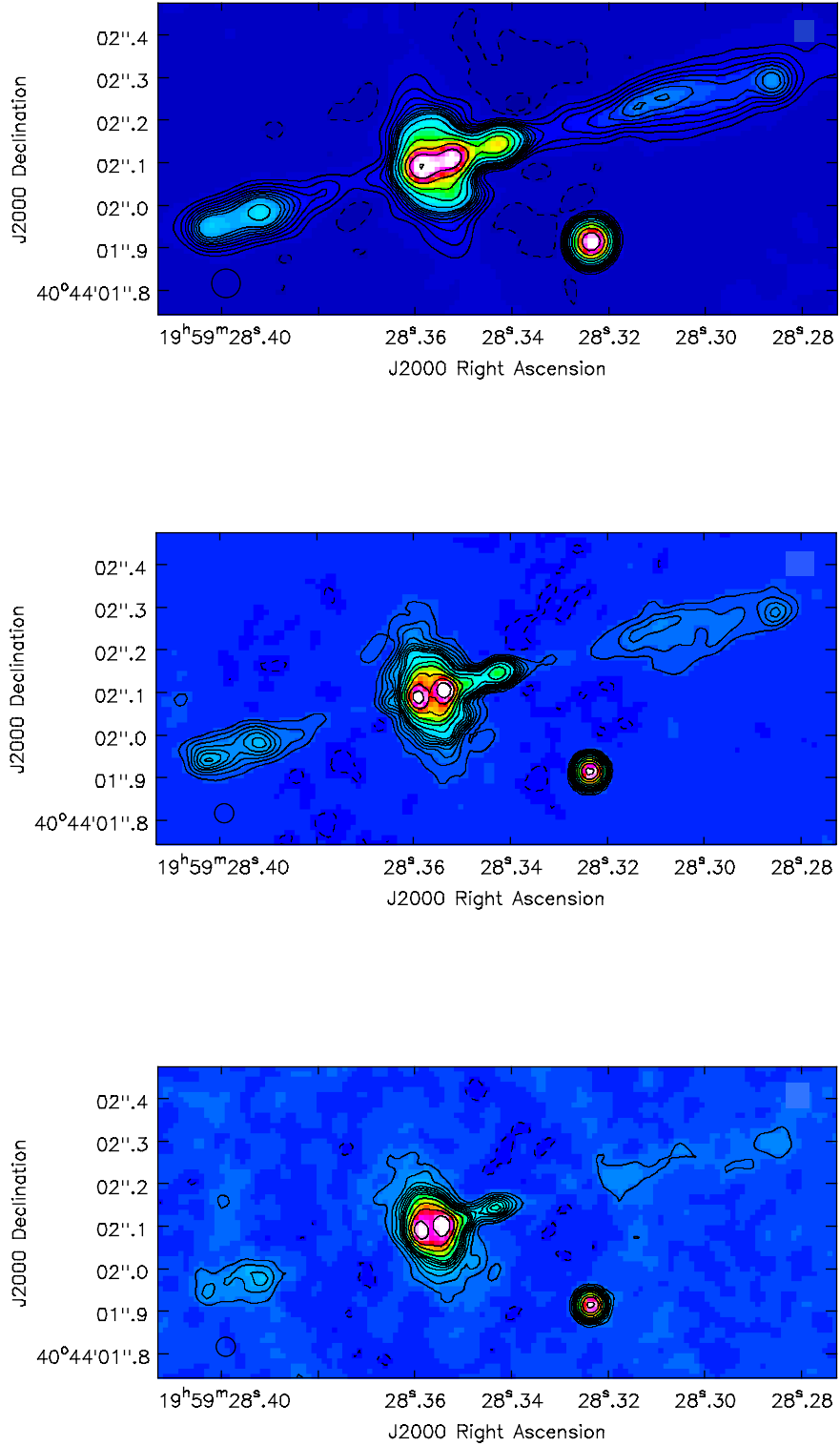}
\caption{VLA images of the nuclear regions of Cygnus A at different
frequencies, after subtraction of a point-source core (see appendix).
Top: 22 GHz, 67 mas resolution and rms noise, $\sigma = 20$ μJy
beam$^{-1}$. Middle: 34 GHz, 45 mas resolution, $\sigma = 12$ μJy beam$^{-1}$. Lower:
44 GHz, 45 mas resolution, $\sigma = 24$ μJy beam$^{-1}$. The contour levels are
-2, -1, 1, 2, 3, 4, 5, 6, 7, 8, 11, 16, 22, 5 45, 64, 91, 128 $\times  4\sigma$.
}
\label{fig:f1}
\end{figure}

\begin{figure}
%\vspace{-3in}
\plotone{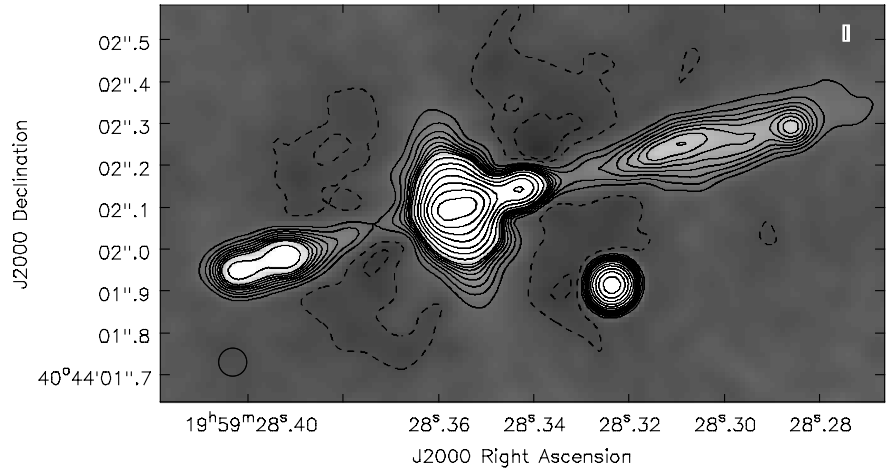}
\caption{Same as figure 1, but now summed over all three bands from 19
GHz to 47 GHz, at 67 mas resolution, $\sigma = 13$ μJy beam$^{-1}$.
}
\label{fig:f2}
\end{figure}

\begin{figure}
%\vspace{-3in}
\plotone{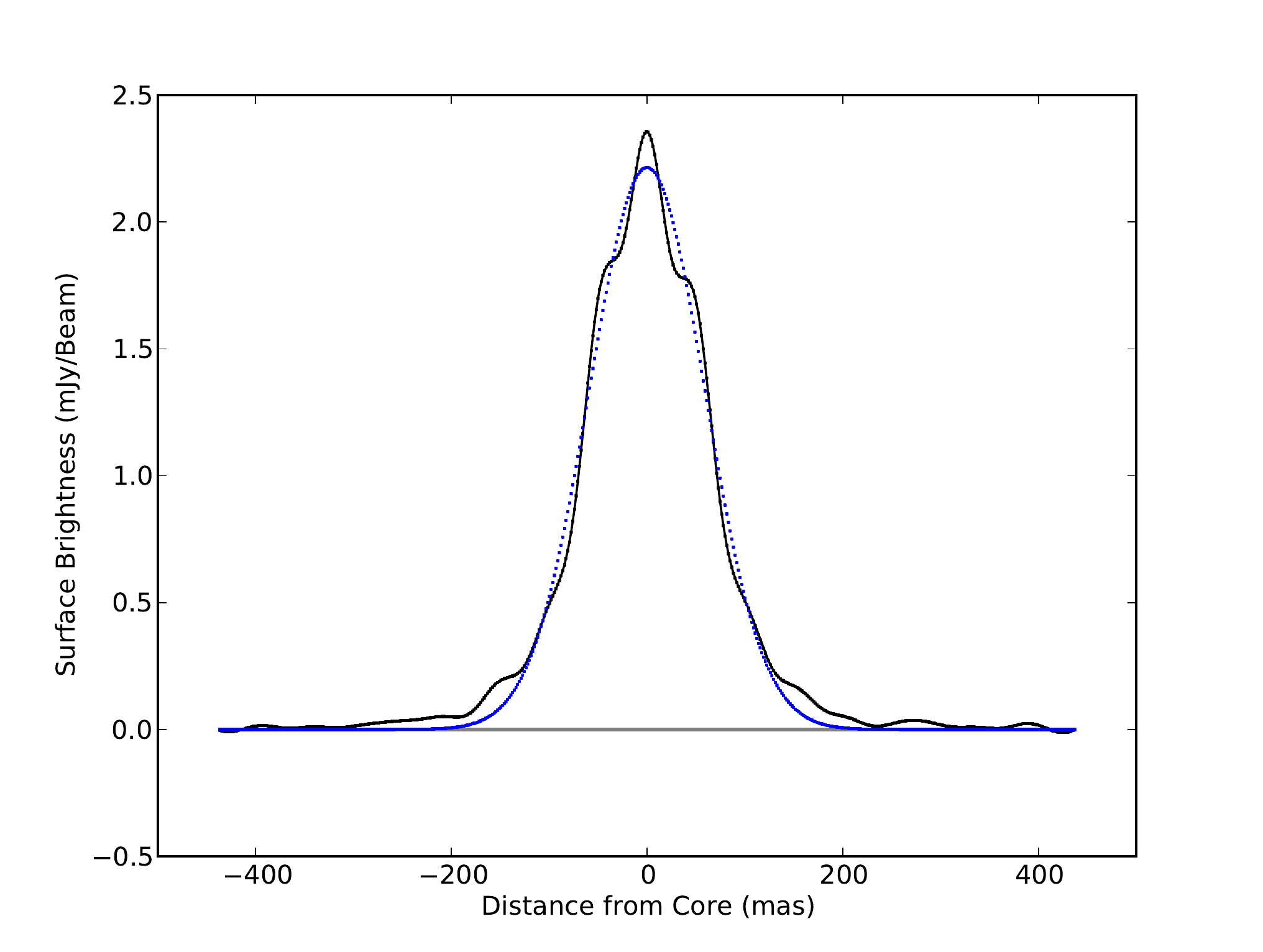}
\caption{A slice in surface brightness through the torus along a line
perpendicular to the jet, through the core position, at
32 GHz and 45 mas resolution (black). The blue
curve is a Gaussian fit to the data, with a model peak of  2.2 mJy,
and a FWHM = 138 mas, centered on the radio core position.
}
\label{fig:f3}
\end{figure}

\begin{figure}
\includegraphics[scale=0.7,angle=-90]{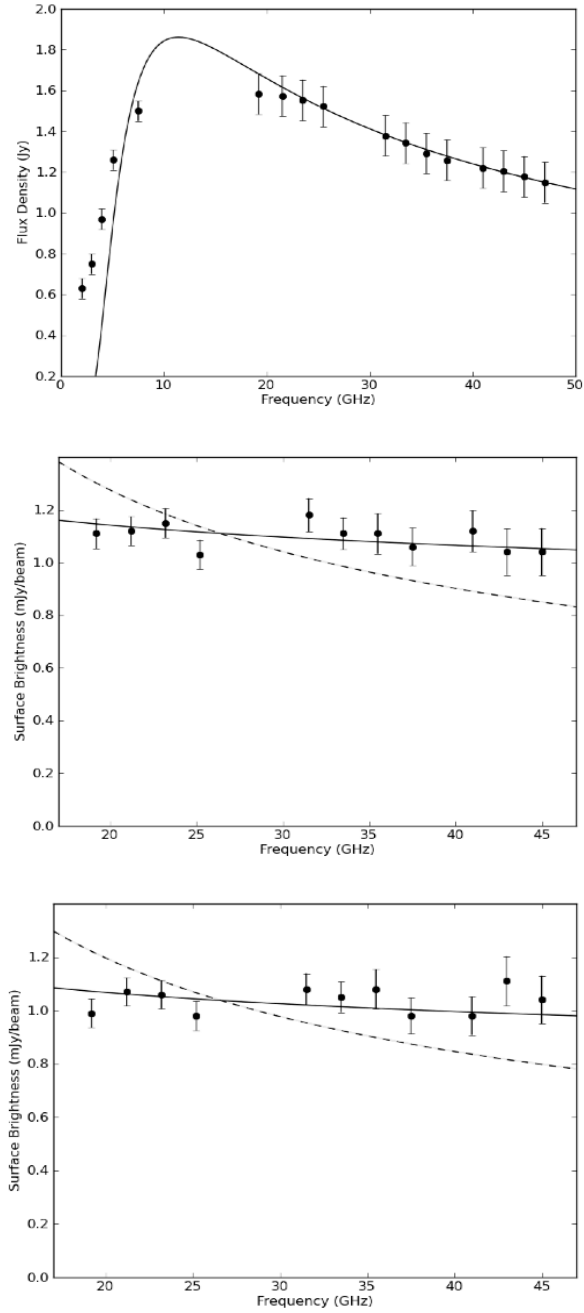}
%\plotone{fig4.pdf}
\caption{Spectra of the radio core, and two positions in the torus of
Cygnus A at 67mas resolution.  The torus positions are separated by
about 100 mas to the north and south of the core position, along the
torus major axis. The Core model is a power-law at high frequency of
index -0.5, determined by fitting to the 8 highest frequency points,
then allowing for Free-Free absorption with an emission measure set
by the surface brightness of the torus itself. The error bars for the
core represent differences in measured flux density for the core at
different epochs, possibly due to variations on timescales of
years. The two models for the torus spectra are: (i) an optically thin
free-free spectral index of -0.1 (solid), and (ii) a power-law
spectrum of index -0.5 (dash), as would  occur due to Thomson scattering of
emission from the core. The error bars for the torus positions are a
quadratic sum of the rms noise in each image, plus estimated 4\% to 8\%
errors due to calibration and imaging, rising with frequency.
}
\label{fig:f4}
\end{figure}

\begin{figure}
%\plotone{fig5.pdf}
\centering 
\includegraphics[scale=0.6]{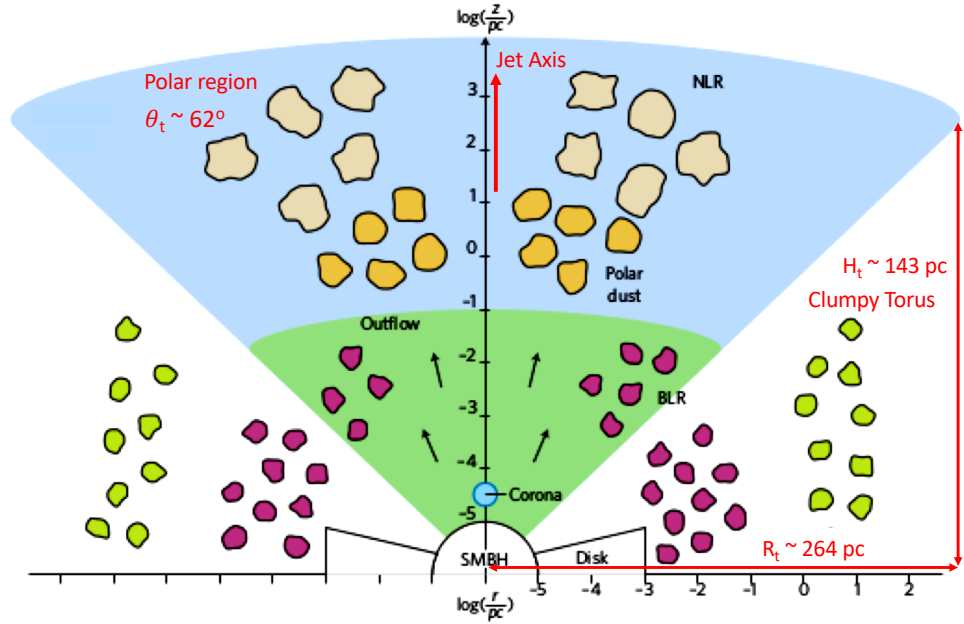}
\caption{Schematic of a thick torus in Cygnus A, adopted and annotated
from \citet{ramos17}, in which the green clumps indicate dust clouds
on a few to hundred-parsec scales in the torus.  The purple clumps are
high ionization clouds on scales $<< 1$ pc, close the SMBH.
The yellow and grey clumps represent
lower ionization clouds and residual dust on 100 pc scales in the lower
density polar regions.  Our line of sight passes through the torus to
the radio core, the inner accretion disk, and broad line
regions. These inner regions are thereby heavily extincted in the optical
and UV by the dust clumps in the denser torus regions. The radio
free-free emission comes from intermediate density gas clumps in the
multi-phase torus.
}
\label{fig:f5}
\end{figure}

\begin{figure}
%\plotone{figA1.pdf}
\centering 
\includegraphics[scale=0.7,angle=-90]{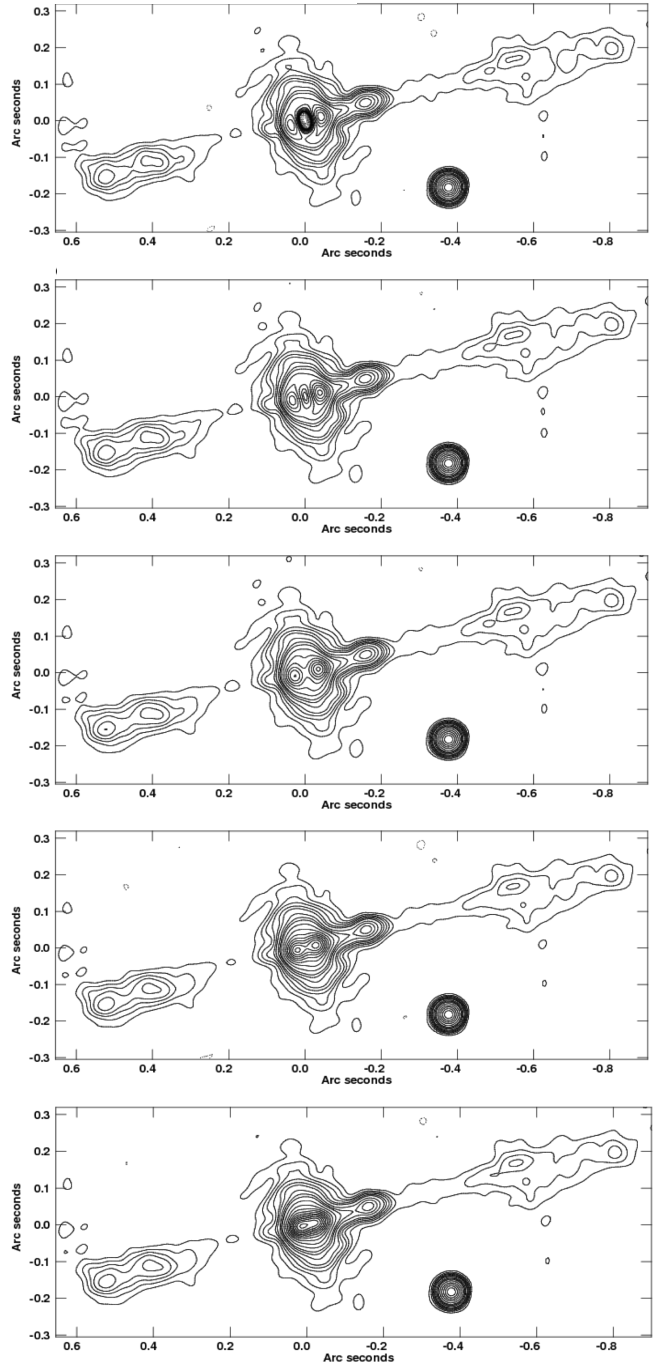}
\caption{{\bf A1}:
Showing the effect on the underlying emission of removing a central
point source from the visibilities at 32GHz.
The center image is the adopted 'best' value for
core subtraction, leading to the flattest profile across the core
region along the torus axis.
Images above the center have larger core values subtracted, by 2
mJy and 4 mJy, respectively. Images below the center have smaller core
values subtracted, by similar amounts.  Brightness contours are .075,
.15, .225, .3, .4, .5, .7, .9, 1.1, 1.5, 2, … 5, 6.5 mJy
beam$^{-1}$. The primary conclusion is that
the nuclear flux removed has essentially no influence on the
brightness of the torus beyond a radius of about 50 mas
from the nucleus.
}
\label{fig:A1}
\end{figure}

\begin{figure}
%\plottwo{figA2a.png, figA2b.png}
%\includegraphics[scale=0.8]{figA2a.png}
%\includegraphics[scale=0.8]{figA2b.png}
\includegraphics[scale=0.7,angle=-90]{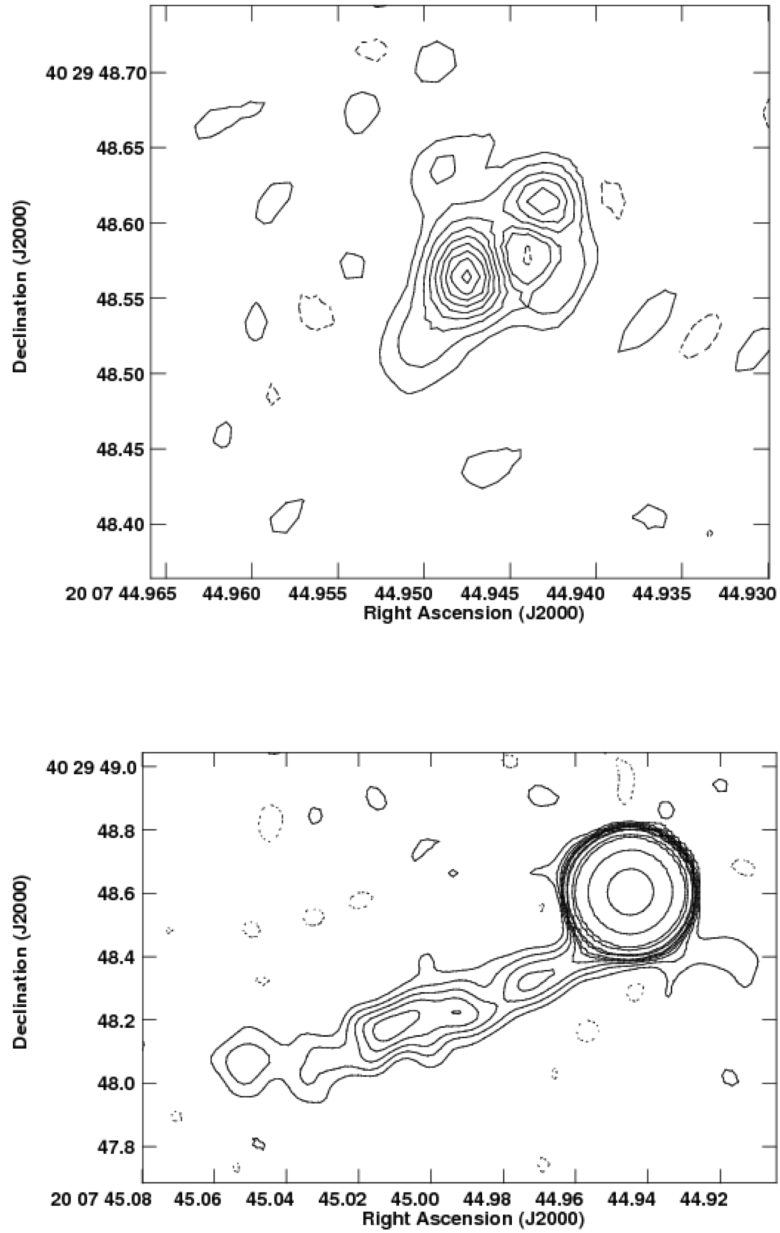}
\caption{{\bf A2}, Top: The residual emission in the calibrator source
J2007+4029 at 33.5 GHz with 42mas resolution, after calibration and
removal of the nuclear emission using a process identical to that
employed for Cygnus A. There is no extended emission on the scale or
brightness of that seen in Cygnus A. The pair of weak sources adjacent
to the nuclear core are from the milliarcsecond-scale jets (42). The
extension to the southwest is the base of the larger-scale jet visible. 
Contours are -0.15, .15, .3, .5, … 1.7 mJy
beam$^{-1}$.  Bottom: The arcsecond-scale jet of the calibrator
J2007+4029, at 9.0 GHz, with 0.15 arcsecond resolution. The outer
jet in this source extends 1.1 arcseconds to the SW of the nucleus. Brightness
contours are -.025, .025, .05, .1, .2, .3, .5, .75, 1, 2, 3, 5, 50,
500 mJy beam$^{-1}$.
}
\label{fig:fA2}
\end{figure}

\end{document}